\documentclass[12pt]{article}

\usepackage{graphicx}
\usepackage{amsfonts,amsmath,amssymb,amsthm}
\usepackage{indentfirst}
\usepackage{cite,hyperref}
\usepackage{mathptmx}

\numberwithin{equation}{section}
\DeclareMathAlphabet{\mathcal}{OMS}{cmsy}{b}{n}

\newcommand{\email}[1]{\footnote{E-mail: \href{mailto:#1}{#1}}}
\begin{document}

\title{An analogy between the Schwarzschild solution in a noncommutative gauge theory and the Reissner-Nordstr\"om metric}
\author{R. Bufalo$^{1,2}$\email{%
rbufalo@ift.unesp.br} ~ and A. Tureanu$^{1}$\email{%
anca.tureanu@helsinki.fi} \\
\textit{$^{1}$ \small Department of Physics, University of Helsinki, P.O.
Box 64}\\
\textit{ \small FI-00014 Helsinki, Finland}\\
\textit{{$^{2}$ \small Instituto de F\'{\i}sica Te\'orica (IFT), Universidade Estadual Paulista}} \\
\textit{\small Rua Dr. Bento Teobaldo Ferraz 271, Bloco II, 01140-070 S\~ao Paulo, SP, Brazil}\\
}
\maketitle

\begin{abstract}
We study modifications of the Schwarzschild solution within the noncommutative gauge theory of gravity. In the present analysis, the deformed solutions are obtained by solving the field equations perturbatively, up to the second order in the noncommutativity parameter $\Theta$, for both exterior and interior solutions of the equations of motion for $e_\mu ^a \left(x\right)$. Remarkably, we find that this new noncommutive solution is analogous to the Reissner-Nordstr\"om solution in the ordinary spacetime, in which the square of electric charge is replaced by the square of the noncommutativity parameter, but with opposite sign. This amounts to the noncommutative Schwarzschild radius $r_{NCS}$  becoming larger than the usual radius $r_S =2M$, instead of smaller as it happens to the Reissner-Nordstr\"om radius $r_{RN}$, implying that $r_{NCS}>r_{S} >r_{RN}$. An intuitive interpretation of this result is mentioned.
\end{abstract}

\section{Introduction}

Over the past decade we have witnessed enormous and laborious advances in noncommutative geometry, reaching an impressive level in its formal development and applicability in the most different areas of high-energy physics, its most appealing feature being in providing a better understanding about the quantum nature of spacetime. The noncommutativity of spacetime, whose structure is determined by $[x^\mu,x^\nu]=i\Theta^{\mu\nu}$, is intrinsically connected with gravity \cite{ref1,ref2,ref32}, and the construction of a consistent theory of gravity on noncommutative spacetime has been attempted in several proposals. The main problem is to formulate properly the concept of invariance under general coordinate transformations in the noncommutative case \cite{ref13}.

Regarding the various investigations and proposals in formulating a noncommutative theory of gravity, we may remark that most of them were defined in the framework of the gauge theory of gravitation \cite{ref33,ref3,ref4,ref5,ref6,ref8,ref35}, in which the Seiberg-Witten map \cite{ref2} was widely explored in such a way as to define and compute the deformed expressions for the vierbein fields and spin connections \cite{ref5}. The gauge theory of gravity was also used in order to define a noncommutative extension of the unimodular theory of gravitation \cite{ref10,ref11}. In another natural approach one may instead consider the twisted Poincar\'e algebra \cite{ref38,ref39}, in order to construct noncommutative gravitational theories \cite{ref9,ref27}.

All these discussions in constructing a consistent noncommutative gravitational theory have led naturally to several studies on noncommutative analogues of black holes. However, in most of the cases, the solutions were not obtained from the field equations. Rather they were obtained under certain noncommutative-inspired guidelines. In particular, a noncommutative-inspired Gaussian mass-distribution as matter source for a black hole solution has been discussed in \cite{ref16,ref19}, as well as its thermodynamical properties \cite{ref20,ref40}. Furthermore, using the Poincar\'e gauge theory combined with a Seiberg-Witten map \cite{ref5}, noncommutative solutions were found for the Schwarzschild black hole \cite{ref14} and for the charged black hole case \cite{ref15}, also for the BTZ black hole \cite{ref28,ref29}, and alternatively by using the noncommutative Riemannian geometry from Ref.~\cite{ref13} a different Schwarzschild black hole solution was discussed in \cite{ref23}.

Prompted by these ambiguous facts and results we are led to address the problem of noncommutative black hole physics once again, but now considering deformed solutions obtained directly by solving the field equations. For this purpose, it is compelling to use the vierbein formalism for gravity \cite{ref30}, since this field is defined in the local Lorentz frame, and physics is more transparent when expressed in a locally inertial frame. This can be appropriately formulated in terms of the gauge theories of gravity \cite{ref50,ref51,ref52}. In fact, we will follow the approach developed in Ref.~\cite{ref5}, where a deformed theory of gravitation was constructed by gauging the noncommutative de Sitter $SO\left(4,1\right)$ group. Afterwards, by contracting the noncommutative de Sitter $SO\left(4,1\right)$ group to the Poincar\'e (inhomogeneous Lorentz) group $ISO\left(3,1\right)$, we obtain the framework in which our explicit calculation will be performed.


In this paper we discuss, within the gauge theory of gravity, a particular case of a four-dimensional static noncommutative spacetime, endowed with a spherically symmetric metric \cite{ref26}, the so-called noncommutative Schwarzschild spacetime. In Section \ref{sec:1} we revise and develop the main steps of the gauge theory of the de Sitter $SO\left(4,1\right)$ group and its contraction to the Poincar\'e group $ISO\left(3,1\right)$ in the present analysis. Afterwards, in Section \ref{sec:2}, we define the star product among the vierbein fields and also discuss the need of a reality condition in the equations of motion. Next, we compute the deformed exterior solution of the vacuum field equations, up to the second-order of the expansion in $\Theta$. In Section \ref{sec:3}, by complementarity, we compute the deformed interior solution, by considering as matter source the stress-energy tensor of a perfect fluid. Moreover, remarkably, the obtained deformed metric is analogous to the well-known Reissner-Nordstr\"om metric \cite{ref31} in the ordinary spacetime, in which the square of noncommutativity parameter plays the part of the square of the electric charge, but with opposite sign. This last fact is further analyzed, and implications are discussed. In Section \ref{sec:4} we summarize the results, and present our final remarks.


\section{De Sitter gauge theory}
\label{sec:1}

We start by reviewing the main ingredients of the orthonormal basis method, or simply vierbein formalism \cite{ref30}. The appropriated framework to introduce this formalism is the gauge theory of the de Sitter group $SO\left(4,1\right)$ for a 4-dimensional spacetime \cite{ref52}. The $SO\left(4,1\right)$ group is $10$-dimensional and its infinitesimal generators are $M_{AB}=-M_{BA}$, $A,B=0,1,2,3,4$. Now, if we put $A=a,4$, $B=b,4$, etc., by introducing the indices $a,b,...=0,1,2,3$, \footnote{Throughout this paper, Greek indices label the spacetime coordinates, whereas Latin indices label the local Lorentz frame.} then we can identify $M_{AB}$ as the generators of translations $P_{a}=M_{a4}$ and Lorentz rotations $M_{ab}=-M_{ba}$. In this framework the corresponding gauge potentials are denoted by $\omega_{\mu}^{AB}=-\omega_{\mu}^{BA}$. Following the reasoning as above, these potentials are identified with the spin connection, $\omega_{\mu}^{ab}=-\omega_{\mu}^{ba}$, and the vierbein fields, $\omega_{\mu}^{a4}= \kappa e_{\mu}^{a}$, in which $\kappa$ is the contraction parameter. Finally, the field strength associated with the gauge potentials $\omega_{\mu}^{AB}$ is
\begin{equation}
F_{\mu\nu}^{AB}=\partial_{\mu}\omega_{\nu}^{AB}-\partial_{\nu}\omega_{\mu}^{AB}
+\eta_{CD}\left(\omega_{\mu}^{AC}\omega_{\nu}^{DB}-\omega_{\nu}^{AC}\omega_{\mu}^{DB}\right),
\end{equation}
where $\eta_{CD}=\text{diag}\left(-1,+1,+1,+1,+1\right)$. Again, we can identify these components as
\begin{align}
F_{\mu\nu}^{a4}\equiv \kappa T_{\mu\nu}^{a}= \kappa \left[\partial_{\mu}e_{\nu}^{a}-\partial_{\nu}e_{\mu}^{a}
+\eta_{bc}\left(\omega_{\mu}^{ab} e_{\nu}^{c}-\omega_{\nu}^{ab}e_{\mu}^{c}\right)\right], \label{eq: 0.1} 
\end{align}
and
\begin{align}
F_{\mu\nu}^{ab}\equiv \mathcal{R}_{\mu\nu}^{ab}=\partial_{\mu}\omega_{\nu}^{ab}-\partial_{\nu}\omega_{\mu}^{ab}
+\eta_{cd}\left(\omega_{\mu}^{ac}\omega_{\nu}^{db}-\omega_{\nu}^{ac}\omega_{\mu}^{db}\right)+\kappa \left(e_{\mu}^{a}e_{\nu}^{b}-e_{\nu}^{a}e_{\mu}^{b}\right), \label{eq: 0.1b}
\end{align}
in which $\eta_{ab}=\text{diag}\left(-1,+1,+1,+1\right)$. For the limit $\kappa \rightarrow 0$ we obtain the $ISO\left(3,1\right)$ gauge group, the so-called Poincar\'e gauge theory of gravitation. This gauge theory has the geometric structure of the Riemann-Cartan space $U_4 $ in which both curvature $\mathcal{R}_{\mu\nu}^{ab}$ and torsion $T_{\mu\nu}^{a}$ are present, these quantities being defined in terms of the gravitational gauge fields $e_{\mu}^{a}$ and potentials $\omega_{\mu}^{ab}$. Hence, one can see that the Poincar\'e gauge theory is an approach to the theory of gravity in which both mass and spin are sources of the gravitational field.

However, if we consider that the spin connections and vierbein fields are not independent variables (e.g., spinless matter), one can solve the spin connection components in terms of the vierbein fields. This is achieved by imposing the condition of null torsion in \eqref{eq: 0.1}, 
\begin{equation}
\partial_{[\mu}e_{\nu]}^{a}=\eta_{cd}e_{[\mu}^{c}\omega_{\nu]}^{ad} . \label{eq: 0.2}
\end{equation}
In this case of vanishing torsion, the geometric structure reduces to the Riemann space $V_4$. Actually, we will use the relation \eqref{eq: 0.2} in order to determine the components of the spin connection in our discussion of noncommutative spacetime, since it appears to be a much simpler and rather natural way to obtain them. Furthermore, for many purposes, tensor calculations are easier accomplished when performed within vierbein formalism; basically, physics is more transparent when expressed in a locally inertial frame. \footnote{It should be emphasized that the Moyal star product is nonlocal by definition and hence it would act in the entire spacetime manifold; however, since we will not consider the full noncommutative contribution but only those contributions up to $\Theta^2$, the situation may simply be seen as small perturbations in the tangent space in which the vierbein fields are defined.}

The curvature tensor of the $ISO\left(3,1\right)$ Poincar\'e gauge theory of gravitation follows from \eqref{eq: 0.1b},
\begin{equation}
\mathcal{R}_{\mu\nu}^{ab}=\partial_{\mu}\omega_{\nu}^{ab}-\partial_{\nu}\omega_{\mu}^{ab}
+\eta_{cd}\left(\omega_{\mu}^{ac}\omega_{\nu}^{db}-\omega_{\nu}^{ac}\omega_{\mu}^{db}\right).\label{eq: 0.0}
\end{equation}

In this paper we are interested in studying a noncommutative counterpart of the Schwarzschild solution from the gravitational equations of motion. For this matter, we will consider the metric for a static and spherically symmetric spacetime, written in the spherical $\left(r,\theta,\phi\right)$ coordinates,
\begin{equation}
ds^{2}=-f\left(r\right)dt^{2}+f^{-1}\left(r\right)dr^{2}+r^{2}\left(d\theta^{2}+\sin^{2}\theta d\phi^{2}\right),\label{eq: 0.5}
\end{equation}
in which we have chosen the spacetime coordinates as $\left(x^{\mu}\right)=\left(t,r,\theta,\phi\right),\ \mu=0,1,2,3$. A convenient orthonormal basis for this metric is given by \footnote{ In general, introducing a noncoordinate orthonormal basis, the vierbein field $e_\mu ^a$ and its inverse $\bar{e}_a ^\mu $ are defined by a covector $e^a = e^a_\mu dx^\mu$ and
its dual vector $\bar{e}_a = \bar{e}_a^\mu \partial_\mu$, respectively, satisfying $e_\mu ^a \bar{e}_a^\nu = \delta _{\mu} ^ {\nu}$ and $e_\mu ^a \bar{e}_b^\mu = \delta _{b} ^ {a}$ \cite{ref52}.}
\begin{gather} 
e_{\mu}^{0}=f^{1/2}\left(dt\right)_{\mu},\quad e_{\mu}^{1}=f^{-1/2}\left(dr\right)_{\mu},\nonumber \\
 e_{\mu}^{2}=r\left(d\theta\right)_{\mu},\quad e_{\mu}^{3}=r\sin\theta\left(d\phi\right)_{\mu}. \label{eq: 0.7}
\end{gather}

In order to determine the unknown component of the metric, we must solve the equations of motion for $e_\mu ^a\left(x\right)$,
\begin{equation}
\mathcal{R}_\mu ^a -\frac{1}{2}e_\mu ^a\mathcal{R}= 8\pi T _\mu ^a,\label{eq: 0.6}
\end{equation}
where $\mathcal{R}_\mu ^a=\mathcal{R}_{\mu\nu}^{ab}\bar{e}_b ^\nu$, $\mathcal{R}=\mathcal{R}_{\mu\nu}^{ab}\bar{e}_a ^\mu \bar{e}_b ^\nu$, with the choice $\hbar = c=G=1$, and $T _\mu ^a$ is the stress-energy tensor of the matter source. While \eqref{eq: 0.6} allows us to determine what is known as the interior solution, we can also consider, alternatively, the exterior solution, which is given by
\begin{equation}
\mathcal{R}_\mu ^a=0.\label{eq: 0.4}
\end{equation}

In what follows,  we shall divide our analysis of the noncommutative solution for the spacetime metric into two parts: first, we consider the exterior and interior solutions, Eqs.\eqref{eq: 0.4} and \eqref{eq: 0.6}, respectively, by considering, as a particular application, the stress-energy tensor of a perfect fluid.


\section{Exterior solution}
\label{sec:2}

In this section we will incorporate noncommutativity into our approach by introducing the Moyal star product $"\star"$ between the vierbein fields, and then proceed to analysing perturbatively the noncommutative contributions for the solution of the vacuum equations of motion for a static and spherically symmetric spacetime. In order to implement noncommutativity in gravity, we can follow the procedure as in Ref.~\cite{ref5}, in which a deformation of gravitation is obtained by gauging the noncommutative de Sitter $SO\left(4,1\right)$ group; afterwards, by contraction $\left(\kappa \rightarrow 0\right)$ to the $ISO\left(3,1\right)$ Poincar\'e gauge theory we obtain the deformed gauge theory in which we perform our calculations.

Besides, we assume that the noncommutative structure of the spacetime is determined by
\begin{equation}
[x^\mu,x^\nu]=i\Theta ^{\mu\nu},
\end{equation}
where $\Theta ^{\mu\nu}=-\Theta ^{\nu\mu}$ are constant parameters. In this case, in order to develop the noncommutative gauge theory, we introduce the Moyal star product between the functions $g$ and $h$ defined as \cite{ref14,ref15}
\begin{equation}
g\left(x\right)\star h\left(x\right)=g\left(x\right)\exp\left[\frac{i}{2}
\Theta ^{\mu\nu} \overleftarrow{\partial}_{\mu}\overrightarrow{\partial}_{\nu}\right]h\left(x\right).\label{eq: 1.1}
\end{equation}
It should be emphasized, nonetheless, that introducing noncommutativity through the star product \eqref{eq: 1.1} makes the theory noncovariant under general coordinate transformations, regardless whether it is used in the gauge theory or Einstein theory of gravity. Moreover, one should also note that  the Lorentz symmetry is spoiled due to noncommutativity as usual, in our case, from the vierbein
fields.

In the following we shall introduce the deformed Einsten-Cartan action \cite{ref33},
\begin{align}
S_{EC}=&\frac{1}{4\chi} \int d^4 x \left(  \left| e _\star \right| \star \bar{e}_a ^\mu  \star \mathcal{R} ^{ab}_{\mu \nu} \star 
 \bar{e}_b ^\nu   +h.c. \right), \\
=&\frac{1}{4\chi} \int d^4 x \left(  \left| e \right|  \mathcal{R}_{\star}  +h.c. \right), \label{eq: 1.15}
\end{align}
where $\chi$ is the Einstein coupling and $\left| e _\star\right| =\det _\star(e^a _\mu) = \frac{1}{4!}\epsilon^{\mu \nu \lambda \sigma}\epsilon_{abcd}~e_\mu ^a \star e_\nu ^b \star e_\lambda ^c \star e_\sigma ^d =\det (e^a _\mu) + \mathcal{O}(\Theta ^2)$. Besides, following from such definition of the star product we have the deformed metric $g_{\mu \nu } = \frac{1}{2}\eta _{ab}\left( e_\mu ^a \star e_\nu ^b + e_\nu ^b \star e_\mu ^a \right)$ and also the deformed Ricci tensor $\mathcal{R} ^{a}_{\mu }  = \mathcal{R} ^{ab}_{\mu \nu} \star 
 \bar{e}_b ^\nu $ and scalar $\mathcal{R}_{\star}= \bar{e}_a ^\mu  \star \mathcal{R} ^{ab}_{\mu \nu} \star 
 \bar{e}_b ^\nu $.  Notice the hermiticity of the above definition, Eq.\eqref{eq: 1.15}.  Such a definition will be very important in what follows in our analysis of the deformed Einstein field equations. 

A remaining quantity to be derived in the deformed gauge theory of gravity is the torsion tensor. This can be obtained through the variation of the deformed Einstein-Cartan action \eqref{eq: 1.15} with respect to the spin connection $\omega_{\mu}^{ab}$. We therefore obtain the expression for the deformed torsion as
\begin{equation}
2T_{\mu\nu}^{a} \equiv 2\partial_{[\mu}e_{\nu]}^{a}-\eta_{cd}e_{[\mu}^{c}\star \omega_{\nu]}^{ad} +h.c. \label{eq: 1.2a}
\end{equation}
In particular, it should be noticed that the reality condition on the torsion tensor is encoded into this expression, therefore no star-ordering ambiguity is present. 

Consider now the case of vanishing torsion, hence, we have a generalized condition written as
\begin{equation}
 2\partial_{[\mu}e_{\nu]}^{a}=\eta_{cd}e_{[\mu}^{c}\star \omega_{\nu]}^{ad} +h.c.  \label{eq: 1.2}
\end{equation}
This expression is the starting point for evaluating the components of the (complex) spin connection, since it allows to write the spin connection in terms of the vierbein fields. We recall that in the NC case the torsion can be nonvanishing, though in the ordinary model it vanishes (see, for example, Ref. \cite{NCRW} for the case of the Robertson--Walker metric), so our vanishing torsion condition is a specific restriction in the analysis.

The derivatives on the left-hand side of \eqref{eq: 1.2} can be readily obtained from \eqref{eq: 0.7}; moreover, substituting the explicit form of the vierbein fields into the right-hand side of \eqref{eq: 1.2}, we find the following four relations to be satisfied:
\begin{align}
f^{-\frac{1}{2}}f^{\prime}\left(dr\right)_{[\mu}\left(dt\right)_{\nu]}  =  \left(f^{-\frac{1}{2}}\left(dr\right)_{[\mu}\right)\star\omega_{\nu]}^{01}+\left(r\left(d\theta\right)_{[\mu}\right)\star
\omega_{\nu]}^{02}+\left(r\sin\theta\left(d\phi\right)_{[\mu}\right)\star\omega_{\nu]}^{03}
+h.c.,\label{eq: 1.3a}\\
0=  -\left(f^{\frac{1}{2}}\left(dt\right)_{[\mu}\right)\star\omega_{\nu]}^{10}+\left(r\left(d\theta\right) _{[\mu}\right)
\star\omega_{\nu]}^{12}+\left(r\sin\theta\left(d\phi\right)_{[\mu}\right)\star\omega_{\nu]}^{13}+h.c. ,\label{eq: 1.3b}\\
2\left(dr\right)_{[\mu}\left(d\theta\right)_{\nu]}  =  -\left(f^{\frac{1}{2}}\left(dt\right)_{[\mu}\right)\star\omega_{\nu]}^{20}+\left(f^{-\frac{1}{2}}\left(dr\right)_{[\mu}\right)
\star\omega_{\nu]}^{21}+\left(r\sin\theta\left(d\phi\right)_{[\mu}\right)\star\omega_{\nu]}^{23}+h.c.,\label{eq: 1.3c}\\
2\sin\theta\left(dr\right)_{[\mu}\left(d\phi\right)_{\nu]}+2r\cos\theta\left(d\theta\right)_{[\mu}
\left(d\phi\right)_{\nu]}  = -\left(f^{\frac{1}{2}}\left(dt\right)_{[\mu}\right)\star\omega_{\nu]}^{30} +\left(f^{-\frac{1}{2}}\left(dr\right)_{[\mu}\right)\star\omega_{\nu]}^{31}\nonumber \\
+\left(r\left(d\theta\right)_{[\mu}\right)
\star\omega_{\nu]}^{32}+h.c. \label{eq: 1.3d}
\end{align}
In order to solve the above relations for the spin connection components we will propose an Ansatz based on the commutative case \cite{ref30}. Hence a plausible Ansatz for the components is
\begin{equation}
\omega_{\mu}^{02}=0, \quad \omega_{\mu}^{03}=0,\quad \omega_{\mu}^{01}=\frac{1}{2}f^{\prime}\left(dt\right)_{\mu},\quad\omega_{\mu}^{12}=-f^{\frac{1}{2}}\left(d\theta\right)_{\mu}. \label{eq: 1.4}
\end{equation}
Now, replacing them back into the Eqs.\eqref{eq: 1.3b} and \eqref{eq: 1.3c}, we find the following constraints on the remaining components:
\begin{align}
\left(r\sin\theta\left(d\phi\right)_{[\mu}\right)\star\omega_{\nu]}^{13}+h.c.=&0,\\
 \left(r\sin\theta\left(d\phi\right)_{[\mu}\right)\star\omega_{\nu]}^{23}+h.c.  =&0,
 \end{align}
respectively. In addition, by taking $\mu=1$ and $\mu=2$ in Eq.\eqref{eq: 1.3d}, we obtain respectively further constraints
\begin{align}
2\sin\theta \left(d\phi\right)_{\nu}  = &\left(f^{-\frac{1}{2}} \right)\star\omega_{\nu}^{31}+h.c.,\\
2r\cos\theta  \left(d\phi\right)_{\nu}  = & \left(r  \right)
\star\omega_{\nu}^{32}+h.c..
 \end{align}
Thus, we observe that these components can be written as
\begin{align}
\omega_{\mu}^{13}  =& -\left(f^{\frac{1}{2}}\right)\star\left(\sin\theta\right)\left(d\phi\right)_{\mu}, \label{eq: 1.5a} \\ 
\omega_{\mu}^{23}  =& -\left(r^{-1}\right)\star\left(r\cos\theta\right)\left(d\phi\right)_{\mu}. \label{eq: 1.5b}
\end{align}
Finally, since we have found no inconsistency (in light of non-deformed solutions as well), we can conclude that our initial guess is, in fact, the deformed solution for the components of the spin connection.
Notice, however, that purely imaginary terms could be added into the spin connection solutions Eqs.\eqref{eq: 1.4}, \eqref{eq: 1.5a} and \eqref{eq: 1.5b}, so that the constraints \eqref{eq: 1.2} are not violated. In addition, the extra pieces may
be chosen so that they vanish in the commutative limit. Hence, the class of solutions that we have determined here is a particular case of a larger group of physically deformed solutions.

The Riemann tensor \eqref{eq: 0.0} may be generalized to the noncommutative case, in a general fashion, by replacing the usual product with the star product \eqref{eq: 1.1}:
\begin{equation}
\mathcal{R}_{\mu\nu}^{ab}=\partial_{\mu}\omega_{\nu}^{ab}-\partial_{\nu}\omega_{\mu}^{ab}
+\eta_{cd}\left(\omega_{\mu}^{ac}\star\omega_{\nu}^{db}-\omega_{\nu}^{ac}\star\omega_{\mu}^{db}\right). \label{eq: 1.6}
\end{equation}
It is rather direct to evaluate the nonvanishing components of the Riemann tensor \eqref{eq: 1.6} by means of the components of the spin connection, Eqs.\eqref{eq: 1.4}, \eqref{eq: 1.5a} and \eqref{eq: 1.5b}. After some straightforward calculation, we get
\begin{align}
\mathcal{R}_{\mu\nu}^{01} & =f^{\prime\prime}\left(dr\right)_{[\mu}\left(dt\right)_{\nu]},\quad 
\mathcal{R}_{\mu\nu}^{02}  =f^{\frac{1}{2}}f^{\prime}\left(d\theta\right)_{[\mu}\left(dt\right)_{\nu]},\label{eq: 1.7a}\\
\mathcal{R}_{\mu\nu}^{03}  &=-\left[f^{\frac{1}{2}}f^{\prime}\right]\star\left(\sin\theta\right)\left(dt\right)_{[\mu}\left(d\phi\right)_{\nu]},\\
\mathcal{R}_{\mu\nu}^{12} &=-f^{-\frac{1}{2}}f^{\prime}\left(dr\right)_{[\mu}\left(d\theta\right)_{\nu]}, \label{eq: 1.7b}
\end{align}
and
\begin{align}
& \mathcal{R}_{\mu\nu}^{13}  =-\left(f^{-\frac{1}{2}}f^{\prime}\right)\star\left(\sin\theta\right)\left(dr\right)_{[\mu}\left(d\phi\right)_{\nu]}\label{eq: 1.7c} \\
&+2\left(  \left[f^{\frac{1}{2}}r^{-1}\right]\star\left(r\cos\theta\right) -f^{\frac{1}{2}}\star \cos\theta \right)
\left(d\theta\right)_{[\mu}\left(d\phi\right)_{\nu]}, \nonumber
\end{align}
and
\begin{align}
& \mathcal{R}_{\mu\nu}^{23}  = 2 \left( \left(r^{-2}\right)\star\left(r\cos\theta\right) - r^{-1} \star\cos\theta \right)  \left(dr\right) _{[\mu}\left(d\phi\right)_{\nu]}\nonumber \\
 & +2\left( \left(r^{-1}\right)\star\left(r\sin\theta\right)  - f \star \sin\theta \right) \left(d\theta\right)_{[\mu}\left(d\phi\right)_{\nu]} . \label{eq: 1.7d}
\end{align}
It is worth emphasizing the presence of extra terms due to the noncommutativity in Eqs. \eqref{eq: 1.7c} and \eqref{eq: 1.7d}, that cancel each other when we take $\Theta=0$. The noncommutative generalization for the Ricci tensor can be readily expressed as before, and we find that the vacuum deformed field equation can be derived from \eqref{eq: 1.15}, 
\begin{equation}
 \mathcal{R}_{\mu}^{a} + h.c.=0.\label{eq: 1.8}
\end{equation}

In general, the perturbative calculations using the star product lead to imaginary parts in the odd powers of the parameter $\Theta$; hence, since we are in a noncommutative gauge theory of gravity, the gauge fields, as well as the equations of motion \eqref{eq: 1.8}, are subjected to reality conditions following naturally from the action \eqref{eq: 1.15}. For this purpose, the nonvanishing Ricci tensor components are properly expressed in the following form:\footnote{Actually, there is another nonvanishing component exclusively due to the noncommutativity, the nondiagonal one: $\mathcal{R}_{2}^{1}=\mathcal{R}_{23}^{13}\star\overline{e}_{3}^{3}$; but, since it does not contribute to the quantities that we are interested in, here we will not present its explicit expression.}
\begin{align}
 \mathcal{R}_{0}^{0} +h.c. & =\mathcal{R}_{0\nu}^{01}\overline{e}_{1}^{\nu}+\mathcal{R}_{0\nu}^{02}\overline{e}_{2}^{\nu}
+\mathcal{R}_{0\nu}^{03}\star\overline{e}_{3}^{\nu}+h.c.,\label{eq: 1.9a}\\
\mathcal{R}_{1}^{1} +h.c. & =\mathcal{R}_{1\nu}^{10}\overline{e}_{0}^{\nu}+\mathcal{R}_{1\nu}^{12}\overline{e}_{2}^{\nu}
+\mathcal{R}_{1\nu}^{13}\star\overline{e}_{3}^{\nu}+h.c.,\label{eq: 1.9b}\\
\mathcal{R}_{2}^{2} +h.c.& =\mathcal{R}_{2\nu}^{20}\overline{e}_{0}^{\nu}+\mathcal{R}_{2\nu}^{21}\overline{e}_{1}^{\nu}
+\mathcal{R}_{2\nu}^{23}\star\overline{e}_{3}^{\nu}+h.c., \label{eq: 1.9c} \\
\mathcal{R}_{3}^{3}+h.c. & =\mathcal{R}_{3\nu}^{30}\star\overline{e}_{0}^{\nu}+\mathcal{R}_{3\nu}^{31}
\star\overline{e}_{1}^{\nu}+\mathcal{R}_{3\nu}^{20}\star\overline{e}_{2}^{\nu}+h.c.. \label{eq: 1.9d}
\end{align}
In order to simplify the calculations, we choose the coordinate system so that the matrix $\Theta^{\mu\nu}$ is given as \cite{ref14,ref15}
\begin{equation}
\Theta^{\mu\nu}= \left(\begin{array}{cccc}
0 & 0 & 0 & 0 \\
0 & 0 & \Theta & 0 \\
0 & -\Theta & 0 & 0 \\
0 & 0 & 0 & 0%
\end{array}%
\right),\quad \mu ,\nu = 0,1,2,3, 
\end{equation}
where $\Theta$ is a constant parameter. The explicit calculations of each of the Ricci tensor components are lengthy, but straightforward, and their expressions, up to the second order in the parameter $\Theta$, are:
\begin{align}
\mathcal{R}_{0}^{0} +h.c.= &-\frac{1}{2}f^{\frac{1}{2}}f^{\prime\prime}-r^{-1}f^{\frac{1}{2}}f^{\prime}-\frac{1}{8r^{3}}\Theta^{2}f^{\frac{1}{2}}f^{\prime}+O\left(\Theta^{3}\right),\label{eq: 1.10a}\\
\mathcal{R}_{1}^{1}  +h.c.= & -\frac{1}{2}f^{-\frac{1}{2}}f^{\prime\prime}-r^{-1}f^{-\frac{1}{2}}f^{\prime}-\frac{1}{8r^{3}}\Theta^{2}f^{-\frac{1}{2}}f^{\prime}+O\left(\Theta^{3}\right),\label{eq: 1.10b}\\
\mathcal{R}_{2}^{2} +h.c.= & -f^{\prime}+r^{-1}\left(1-f\right)+\frac{1}{4r^{3}}\Theta^{2}\left(1-f\right)+O\left(\Theta^{3}\right),\label{eq: 1.10c}\\
\mathcal{R}_{3}^{3} +h.c. = & -f^{\prime}\sin\theta+\left(1-f\right)r^{-1}\sin\theta +\frac{1}{4r^{3}}\Theta^{2} \left(1-f\right)\sin\theta+O\left(\Theta^{3}\right). \label{eq: 1.10d}
\end{align}
Finally, we can evaluate the exterior solution by means of the component of the Ricci tensor $\mathcal{R}_{2}^{2}$. Thus, substituting \eqref{eq: 1.10c} into the equation of motion \eqref{eq: 1.8},
\begin{align}
\mathcal{R}_{2}^{2}+h.c.=&-f^{\prime}+r^{-1}\left(1-f\right)+\frac{1}{4}\Theta^{2}r^{-3}\left(1-f\right)+O\left(\Theta^{3}\right)=0. \label{eq: 1.11}
\end{align}
Solving this equation, we find the deformed exterior solution:
\begin{equation}
f\left(r\right)=1-\frac{C}{r}\left[1-\frac{\Theta^{2}}{8r^{2}}\right]^{-1}, \label{eq: 1.12}
\end{equation}
where $C$ is an integration constant. In the ordinary case, the constant $C$ is related to the total mass $M$ of the Schwarzschild black hole, usually obtained by a direct comparison of the behaviour of a test body in the weak field regime $\left( r\rightarrow\infty\right)$, with the behaviour of a test body in the Newtwonian theory of gravity. However, since the geodesics of the Schwarzschild metric in the noncommutative spacetime are more complicated \cite{ref37}, such a relation does not hold. Hence, we will reserve our comments and implications of the structure of the solution \eqref{eq: 1.12} for the discussion of the deformed interior solution in the Section \ref{sec:3}.

\section{Interior solution}
\label{sec:3}

In order to complement our analysis, after having evaluated the exterior solution, we can compute furthermore the interior solution considering a stress-energy tensor for a (commutative) perfect fluid as the matter source in the equations of motion \eqref{eq: 0.6}. For this purpose, we first remark that in order to ensure the reality of the outcome, we shall consider the action \eqref{eq: 1.15} added with matter fields.
Hence, the deformed interior solution can now be properly obtained from the following expression of the deformed field equations:
\begin{equation}
\mathcal{G}_{\mu}^{a}\equiv \mathcal{R}_{\mu}^{a} -\frac{1}{2}e_{\mu}^{a} \mathcal{R}_\star +h.c.=16\pi T_{\mu}^{a}. \label{eq: 2.2}
\end{equation}
In this way, since we have evaluated the Ricci tensor components in the previous section, we are only left to compute
\begin{align}
  e_{\mu}^{a} \mathcal{R}_\star+h.c.= & \Big( \overline{e}_{0}^{0}\star \mathcal{R}_{0}^{0}+\overline{e}_{1}^{1}\star \mathcal{R}_{1}^{1}  +\overline{e}_{2}^{2}\star\mathcal{R}_{2}^{2}+\overline{e}_{3}^{3}\star\mathcal{R}_{3}^{3} \Big)  e_{\mu}^{a}+h.c. \label{eq: 2.3}
\end{align}
However, it should be emphasized that the Ricci tensor components present in the expression \eqref{eq: 2.2} are those evaluated in Eqs.\eqref{eq: 1.9a}-\eqref{eq: 1.9d}. We are interested in solving, in particular, the $00$ component of the field equation \eqref{eq: 2.2}, so it suffices for our purpose to consider and evaluate
\begin{align}
 e_{0}^{0} \mathcal{R}_\star+h.c. =&\Big( \overline{e}_{0}^{0}\star \mathcal{R}_{0}^{0}+\overline{e}_{1}^{1}\star \mathcal{R}_{1}^{1}+\overline{e}_{2}^{2}\star\mathcal{R}_{2}^{2}+\overline{e}_{3}^{3}\star\mathcal{R}_{3}^{3} \Big) e_{0}^{0}+h.c. \label{eq: 2.4}
\end{align}
Finally, following the same procedure as developed in the previous section and after a lengthy calculation, we find the  result
\begin{align}
\mathcal{G}_{0}^{0}  =-f^{\frac{1}{2}}r^{-2}\left[r\left(1-f\right)\left[1-\frac{1}{8}\Theta^{2}r^{-2}\right]\right]^{\prime}+O\left(\Theta^{3}\right). \label{eq: 2.5}
\end{align}

To illustrate our result, let us consider a stress-energy tensor for a (commutative) perfect fluid, in which $T_{\nu}^{a}=T_{\nu}^{\mu}e_{\mu}^{a}$,
\begin{equation}
T_{\nu}^{\mu}=\text{diag}\left(-\rho,p,p,p\right). \label{eq: 2.6}
\end{equation}
Hence, replacing Eqs. \eqref{eq: 2.5} and \eqref{eq: 2.6} back into the equation \eqref{eq: 2.2}, and then integrating the resulting expression, we get
\begin{equation}
r\left(1-f\right)\left[1-\frac{1}{8}\Theta^{2}r^{-2}\right]=2m\left(r\right)=8\pi \int_{0}^{r}dR R^{2}\rho\left(R\right)+C, \label{eq: 2.7}
\end{equation}
where $m\left(r\right)$ is called the \emph{mass function}. Moreover, we can define conveniently the following quantities: $\Delta=r^{2}-\frac{\Theta^{2}}{8}-2r m\left(r\right)$ and $\Sigma=r^{2}-\frac{\Theta^{2}}{8}$, in such a way that we can cast our solution in the form $f\left(r\right)=\frac{\Delta}{\Sigma}$. This implies that the deformed line element has the following form:
\begin{equation}
ds^{2}=-\frac{\Delta}{\Sigma}dt^{2}+\frac{\Sigma}{\Delta}dr^{2}+r^{2}\left(d\theta^{2}+\sin^{2}\theta d\phi^{2}\right). \label{eq: 2.8}
\end{equation}
Though the presence of a noncommutative contribution has smeared the usual the Schwarzschild singularity $\left(r_S=2M\right)$ in a non-trivial way, we can easily see that $\Delta$ still has one singularity in the $r$-coordinate,
\begin{equation}
r_{+}=m\left(r\right)+ \sqrt{m^{2}\left(r\right)+\frac{\Theta^{2}}{8}}. \label{eq: 2.9}
\end{equation}
In fact, the inner horizon $r=r_{-}$  appears at negative radius, $r=r_{-}<0$, which is meaningless. Hence, we have only one singularity present in this noncommutative case, at $r=r_{+}$.

A direct implication of the noncommutative effects in \eqref{eq: 2.9} can be obtained if we restore the units in the Einstein field equations, in this way
\begin{equation}
r_{+}=\frac{l_{P}}{M_{P}}m\left(r\right)+\sqrt{\left(\frac{l_{P}}{M_{P}}m\left(r\right)\right)^{2}+\frac{\Theta^{2}}{8}},
\end{equation}
with the definitions for the Planck mass $M_{P}=\sqrt{\frac{\hbar c}{G}}\sim 10^{-8}~\text{kg}$,
and Planck length $l_{P}=\sqrt{\frac{\hbar G}{c^{3}}}\sim 10^{-35}~\text{m}$. Besides, we see that the leading correction to the Schwarzschild radius is given in the form
\begin{equation}
r_{+}\simeq \frac{2l_{eff}}{M_{P}}m\left(r\right),
\end{equation}
where the effective minimum length is $l_{eff}=l_{P}\left(1+\left(\frac{M_{P}}{l_{P}}\right)^{2}\frac{\Theta^{2}}{32m^{2}\left(r\right)}\right)$. A simple estimative is found
if we consider a Planck mass black hole, $m\left(r\right)\sim M_{P}$,
then the effective length is of order of $l_{eff}\simeq l_{P}+\frac{\Theta^{2}}{32l_{P}}$.

In particular, noncommutativity is believed to be relevant at Planck scale, therefore the noncommutativity scale can be in principle taken to be the Planck scale, $\Lambda _{NC} = E_P \sim 10^{16}~\text{TeV}$. In that case, one can see that for a Planck mass black hole one obtains the main contribution from the original Schwarzschild solution, and the noncommutative correction is 32 times smaller. The effective black hole radius is therefore slightly larger than Schwarzschild, $l_{eff}\approx 1.03 l_P$.

In addition, one may consider, by means of illustration, intermediary noncommutativity (energy) scales commonly found in literature, because the noncommutative effects may change considerably; these are lower energy bounds following from distinct characteristic energy scales $(E_c)$. A high-energy bound, $\Lambda _{NC} \gtrsim 10^{4}~\text{TeV}$, obtained by analyzing corrections to the electron anomalous magnetic moment $(E_c \sim  \text{TeV})$ \cite{ref45}, results in an effective length of a order of $l_{eff} \lesssim 10^{-12}~\text{m}$; while, for a low-energy bound, $\Lambda _{NC} \gtrsim 10^{11}~\text{TeV}$, following from an analysis of an atomic magnetometer experiment $(E_c \sim \text{eV})$ \cite{ref46}, we obtain an effective length of a order of $l_{eff} \lesssim 10^{-26}~\text{m}$. Hence, we see that the $\Theta$-contribution, when evaluated with lower energy bounds, gives enormously larger radius in view of the usual Planck length, $l_{P}\sim 10^{-35}~\text{m}$. This contrasting behaviour can be traced back to the fact that those lower energy bounds for the noncommutativity are strongly dependent on the type of physics examined and on the precision of the experimental results, where the noncommutative corrections are fitted to the error bars of the experimental data.

Nevertheless, surprisingly enough, the outer horizon in \eqref{eq: 2.9} is analogous to the one obtained in the Reissner-Nordstr\"om stationary metric \cite{ref31} (a charged generalization of the Schwarzschild solution), where $\Delta=r^{2}+Q^2-2r M$ and $\Sigma=r^{2}$, in which the Reissner-Nordstr\"om singularities are given by
\begin{equation}
r_{\pm}=M\pm\sqrt{M^{2}-Q^{2}}, \label{eq: 2.10}
\end{equation}
where $Q$ is the total electric charge of the spacetime and $M$ is the total mass. Here we have the presence of an inner $r=r_{-}$ and outer $r=r_{+}$ horizons.

Hence, since the Reissner-Nordstr\"om metric is obtained in the presence of an electric field, it is reasonable to draw a parallel between the results \eqref{eq: 2.9} and \eqref{eq: 2.10} for the outer horizons, and argue that the deformed $\Theta$-contribution in \eqref{eq: 2.9} plays the part of a background field. Actually, this is a reasonable picture since the interplay between noncommutative coordinates and a background field is often encountered.

However, on the other side, noncommutativity has the opposite effect than the electric field on the singularity, i.e., we can see in \eqref{eq: 2.10} that the presence of an electric charge leads to the decrease in the radius size of the singularity in comparison with the Schwarzschild radius $r_S$, while the noncommutative contribution in \eqref{eq: 2.9} leads to the increase in the radius size. Furthermore, one can physically depict this situation in the following way: in this scenario the electric field and noncommutativity may be seen, respectively, as an attractive and as a repulsive potential/force, making the black hole radius to become smaller and larger, respectively. The latter could be interpreted as fuzziness of spacetime due to noncommutativity, leading to an effect similar to an incompressible fluid. \footnote{The same observation has been made by Yoichiro Nambu in the description of an incompressible liquid (private communication, see also \cite{ref24,ref7,ref12}.)}


\section{Concluding remarks}

\label{sec:4}

In this paper we have determined a new Schwarzschild-type solution in the framework of a noncommutative gauge theory of gravity. Since most of the previous analyses on noncommutative analogues of black holes led to ambiguous facts and results, our main aim in this paper was to address this subject by solving the deformed field equations, which provides a better way in discussing the outcome of the theory. For this purpose, the de Sitter gauge theory of gravitation provided the appropriated framework. In fact, we have followed the construction outlined in Ref.~\cite{ref5}, in which a deformation of the gravitational field has been constructed by gauging the noncommutative de Sitter $SO\left(4,1\right)$ group, and its deformed solutions were obtained by contraction of the noncommutative gauge group $SO\left(4,1\right)$ to the Poincar\'e (inhomogeneous Lorentz) group $ISO\left(3,1\right)$.

However, it should be clear that introducing noncommutativity in a gravitational theory is problematic if formulated either as a gauge theory or as a Einstein theory of gravity, since general covariance is lost due to the use of usual derivatives in the Moyal star product. On the other hand, if one uses a star product with covariant derivatives in order to preserve the diffeomorphism invariance, the products would not be associative. 

Our analysis consisted in studying perturbatively, up to the second order in the noncommutative parameter $\Theta$, solutions of the deformed field equations obtained from the gauge theory.
We have found by solving these deformed gravitational field equations that the noncommutativity smears the (Schwarzschild black hole) singularity in the expression of the deformed metric in a non-trivial way. This is in direct contrast with previous studies, in which some analyses have provided deformed modifications but no changes in the singularity ($r_S=2M$) \cite{ref14,ref15}. The solutions we have considered can be generalized by suitably adding purely imaginary parts to the spin connection solutions Eqs.\eqref{eq: 1.4}, \eqref{eq: 1.5a} and \eqref{eq: 1.5b}, so that the constraints \eqref{eq: 1.2} are not violated and the commutative limit is preserved. It would be interesting to see whether the larger class of solutions would lead to  essentially different physical results. We postpone this study to a future work.

The novel class of deformed solution obtained in this paper has an outer horizon expression analogous to the one from the ordinary Reissner-Nordstr\"om solution. Despite the analogy, the noncommutativity and electric charge contributions have a completely opposite effect on the outer horizon, by making the black hole radius size to increase and decrease, respectively. A similar analogy between the noncommutative Schwarzschild black hole and the Reissner-Nordstr\"om black hole was also found in Ref.~\cite{ref40} when analyzing their thermodynamical behaviour in the near extremal limit (e.g., $M\rightarrow Q$ or $r_+ \rightarrow Q$ in Eq.\eqref{eq: 2.10}). However, this thermal study was performed in an ordinary spacetime, by using a Gaussian mass distribution as a matter source, instead of obtaining a black hole solution from a noncommutative spacetime as we have considered in this paper.

Finally, one may check that the solution \eqref{eq: 2.8} does not satisfy the deformed Einstein equations in metric formalism, differing by $\Theta^2$ terms. This could indicate that, while the use of vierbein or metric formalism in the ordinary case are equivalent and lead to identical results, in the noncommutive case they need not be equivalent. Those aspects are currently under scrutiny.


\subsection*{Acknowledgments}

We are indebted to M. Chaichian for his interest in this work and for discussions. We would like to thank M. Oksanen, P. Pre\v{s}najder and G. Zet for suggestions on the manuscript. R.B. thanks FAPESP for full support, Project No. 2013/26571-4. The support of the  Academy of Finland under
the Projects No. 136539 and 272919 is gratefully acknowledged.

\end{document}